\documentclass[reprint,aps,prx,superscriptaddress,longbibliography]{revtex4-2}

\pdfoutput=1 

\usepackage{hyperref}
\hypersetup{colorlinks=true, citecolor=black, urlcolor=black, linkcolor=black}

\usepackage{graphicx}% Include figure files
\usepackage{dcolumn}% Align table columns on decimal point
\usepackage{braket}
\usepackage{bm}
\usepackage{color}
\usepackage{amsmath}

\begin{document}
%\preprint{}

\title{Ultrafast energy exchange between two single Rydberg atoms \\ on the nanosecond timescale}


\author{Y. Chew}
\affiliation{Institute for Molecular Science, National Institutes of Natural Sciences, Okazaki 444-8585, Japan}
\affiliation{SOKENDAI (The Graduate University for Advanced Studies), Okazaki 444-8585, Japan}

\author{T. Tomita}
\affiliation{Institute for Molecular Science, National Institutes of Natural Sciences, Okazaki 444-8585, Japan}
%\affiliation{SOKENDAI (The Graduate University for Advanced Studies), Okazaki 444-8585, Japan}

\author{T. P. Mahesh}
\affiliation{Institute for Molecular Science, National Institutes of Natural Sciences, Okazaki 444-8585, Japan}
\affiliation{SOKENDAI (The Graduate University for Advanced Studies), Okazaki 444-8585, Japan}

\author{S. Sugawa}
\affiliation{Institute for Molecular Science, National Institutes of Natural Sciences, Okazaki 444-8585, Japan}
\affiliation{SOKENDAI (The Graduate University for Advanced Studies), Okazaki 444-8585, Japan}

\author{S. de L\'es\'eleuc}
\email[]{sylvain@ims.ac.jp}
\affiliation{Institute for Molecular Science, National Institutes of Natural Sciences, Okazaki 444-8585, Japan}
\affiliation{SOKENDAI (The Graduate University for Advanced Studies), Okazaki 444-8585, Japan}

\author{K. Ohmori}
\email[]{ohmori@ims.ac.jp}
\affiliation{Institute for Molecular Science, National Institutes of Natural Sciences, Okazaki 444-8585, Japan}
\affiliation{SOKENDAI (The Graduate University for Advanced Studies), Okazaki 444-8585, Japan}

%\thanks{}

\date{\today}

\begin{abstract}
	%NatPhys like abstract
	Rydberg atoms, with their giant electronic orbitals, exhibit dipole-dipole interaction reaching the GHz range at a distance of a micron, making them a prominent contender for realizing quantum operations well within their coherence time. However, such strong interactions have never been harnessed so far, mainly because of the stringent requirements on the fluctuation of the atom positions and the necessary excitation strength.
	Here, using atoms trapped in the motional ground-state of optical tweezers and excited to a Rydberg state with picosecond pulsed lasers, we observe an interaction-driven energy exchange, i.e., a F\"orster oscilation, occuring in a timescale of nanoseconds, two orders of magnitude faster than in any previous work with Rydberg atoms. 	
	This ultrafast coherent dynamics gives rise to a conditional phase which is the key resource for an ultrafast controlled-$Z$ gate.
	This opens the path for quantum simulation and computation operating at the speed-limit set by dipole-dipole interactions with this ultrafast Rydberg platform.
\end{abstract}

%
%\begin{abstract}
%%PRL like abstract
%We report on the observation of interaction between two Rydberg atoms in a regime two orders of magnitude stronger than realized so far. 
%By trapping atoms in the motional ground-state of optical tweezers separated by as short as $1.5 \, \mu$m, we reach dipole-dipole interaction strengths in the GHz range for Rydberg states $n \sim 40$. 
%Using picosecond pulsed lasers, we excite the state $43D$, where a natural resonance between Rydberg orbitals gives rise to an interaction-driven energy exchange, i.e., a F\"orster oscillation, occuring in a timescale of nanoseconds. 
%We observe the dynamics of the Rydberg population, and by performing Ramsey interferometry, the signature of a conditional phase, which is the resource for an ultrafast $C_Z$ gate.
%This opens the path for quantum simulation and computation operating at the speed-limit set by dipole-dipole interaction with this ultrafast Rydberg platform.
%\end{abstract}

\maketitle

\section{Introduction}

Progress in the field of quantum simulation and computation is fueled by efforts made on a variety of platforms (e.g., superconducting (SC) qubits, quantum dots, trapped ions, neutral atoms, ...) to reach a critical fidelity of quantum operations. This requires to operate on these systems orders of magnitude faster than the timescale set by the coupling to the environment. There is thus a continuous strive to better insulate qubits~\cite{Dzurak2014,Nakamura2020,Sayrin2020,Wang2021,Somoroff2021} and to design faster quantum operations~\cite{Monroe2010,Tarucha2016,Jaewook2018,Hennrich2020}.
Among the latter, a critical operation is the entanglement of two qubits, which requires a time lower-bounded by a speed limit $t_{J} = \pi /J$ set by a platform-dependent interaction strength $J$, e.g., proportional to a capacitance between two SC qubits~\cite{GoogleSC2019}. While gates were often first realized in an adiabatic regime $t \gg t_J$~\cite{Molmer1999,DiCarlo2009,Browaeys2010,Saffman2010} to minimize couplings to unwanted states or degrees of freedom (d.o.f.)~giving unitary errors, entanglement protocols saturating the bound while dealing with these parasitic couplings are highly sought after as they minimize decoherence (non-unitary errors). Devising and realizing such protocols are the subject of intense efforts on all platforms~\cite{Zoller2003,Zoller2006,Monroe2017,Lucas2018,Blatt2021,Martinis2014,DiCarlo2019,GoogleSC2019,GoogleSC2020,DiCarlo2021}.

Arrays of Rydberg atoms in optical tweezers are one of the most exciting systems for quantum simulation~\cite{deLeseleuc2020,Scholl2021,Ebadi2021,Bluvstein2021} and computation~\cite{Levine2019,Saffman2019,Endres2020}.
At its core, the dipole-dipole interaction $\hat{H}_{\rm dip} \sim  \hat{d}_1 \hat{d}_2/4\pi \epsilon_0 R^3$~\cite{SaffmanRMP,Browaeys2020} is used to operate entanglement between two neutral atoms separated by a microscopic distance $R$ thanks to the large matrix elements of the dipole operator $\hat{d}$ ($\sim 1000 \, e a_0$). For two Rydberg atoms with principal quantum number $n = 40$, distant by $R \sim 1 \, \mu$m (to trap them in independent tweezers), the coupling strength between pairs of orbitals reaches $J = \langle r' r'' \vert \hat{H}_{\rm dip} \vert rr \rangle \sim 2 \pi \times 1$~GHz, setting the speed limit of entanglement at $t_J \sim 1$~ns, five orders of magnitude faster than the $100 \, \mu$s radiative lifetime of Rydberg states, as shown in Fig.~\ref{fig1}(a).
Currently, the best entangling protocols of Rydberg atoms~\cite{Levine2019,Saffman2019,Endres2020} are performed deeply in the adiabatic regime compared to the available interaction strength, with a typical duration of $0.5 \, \mu{\rm s}$, as they rely on Rydberg blockade~\cite{Jaksch2000}. 
In this scenario, the interaction strength, --- either the dipole-dipole coupling $J$, or more often the weaker second-order van der Waals shift $V$ ---, is larger than the coupling $\Omega_{\rm cw}$ from ground to Rydberg states with continuous-wave (cw) lasers, such that excitation of two atoms is prohibited. This gives rise to entanglement in a timescale $\pi/\Omega_{\rm cw} \gg t_J$, with a gate speed technically limited by the available power of cw lasers.
This approach is motivated by the suppression of population of the strongly interacting state $\vert rr \rangle$, thus avoiding (i) leakage into other Rydberg states $\ket{r'r''}$, and (ii) motional coupling to the atoms' external d.o.f.~($\hat{x},\hat{p}$) caused by the distance-dependent dipole-dipole interaction. 
Interestingly, other communities have learned to minimize exactly these effects (e.g., leakage to non-computational states in SC qubits~\cite{Martinis2014}, motional coupling in ion traps~\cite{Zoller2003}), while approaching the entanglement speed limit~\cite{Monroe2017,Lucas2018,DiCarlo2019,GoogleSC2019,GoogleSC2020,DiCarlo2021}.
On the other hand, the ``slow'' blockade gate requires to spend a relatively long time in Rydberg states affected by: finite lifetime, black-body radiation~\cite{festa2021blackbody}, Doppler effect, laser phase noise~\cite{deLeseleuc2018,Levine2018}, electric field fluctuation, and motional coupling due to different trapping potentials of Rydberg atoms~\cite{Barredo2020,wilson2019trapped}.

Here, we take a first step into achieving an entanglement protocol with Rydberg atoms at the speed limit set by the dipole-dipole interaction. 
Two single $^{87}$Rb atoms separated by a distance as small as 1.5~$\mu$m are both excited to a $nD$ Rydberg state by using pulsed lasers with a duration of tens of picoseconds~\cite{Takei2016,Mizoguchi2020}. This is faster than the timescale of interaction $J$ but slow enough to resolve the Rydberg orbitals whose splitting $(\Delta E_{n}/\Delta n)^{-1}$ is 10 ps at $n = 40$. 
A natural resonance between the dipole-dipole coupled states $\left \vert 43 D, 43 D \right \rangle$ and $\left \vert 45 P, 41 F \right \rangle$ gives rise to a F\"orster oscillation~\cite{Ravets2014,Ravets2015}, ideally imprinting a conditional $\pi$-phase shift $ \vert 43 D, 43 D\rangle \rightarrow e^{i \pi} \vert 43 D, 43 D \rangle$ after a time $t = \pi/J$: the key resource for a controlled-$Z$ ($C_Z$) gate~\cite{Levine2019,Saffman2019,DiCarlo2019,GoogleSC2019,GoogleSC2020,DiCarlo2021}, one of the two-qubit gates allowing quantum computation.
%We observe this ultrafast dynamics through the p and probe the conditional phase by Ramsey
We first observe this ultrafast dynamics through the population of the $D$ state, and then probe the conditional phase $\phi$ with Ramsey interferometry, obtaining excellent agreement with a calculated $\phi = 1.17 \, \pi$ when taking into account perturbing channels. 

\section{Ultrafast Rydberg platform}
%\paragraph{Our work}
\begin{figure}
	\includegraphics[width=\linewidth]{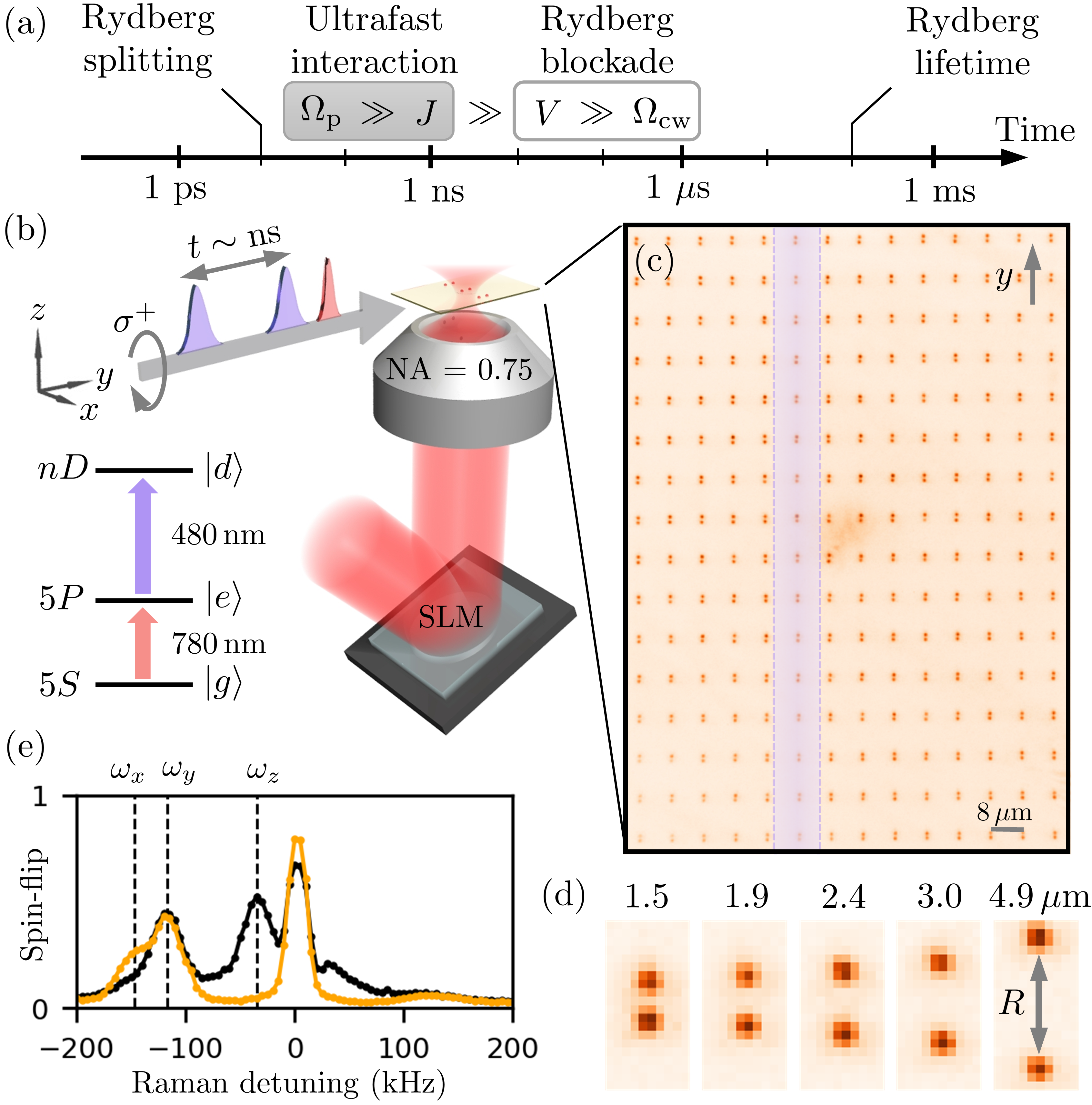}
	\caption{\label{fig1}  \label{fig1a} \textbf{Ultrafast Rydberg platform.} (a) Rydberg physics timescale. $\Omega_{\rm cw}$ and $\Omega_{\rm p}$ are the typical laser couplings to a Rydberg state with continuous-wave and pulsed lasers. $J$ is the dipole-dipole coupling. $V$ is the van der Waals shift. (b) Experimental setup: the trapping beam, diffracted by the SLM, forms a tweezers array at the focus of the objective (NA $= 0.75$). The 780 and 480 nm picosecond laser pulses are shone along the quantization axis ($y$) to excite the atoms to a Rydberg state. (c) Averaged fluorescence image of the atomic array containing $14 \times 16$ pairs of atoms separated by 8 (10)~$\mu$m along $x$ ($y$). All pairs are aligned along the quantization axis. The Rydberg experiments are performed with the column highlighted in blue, where the intensity of the 480 nm pulse laser is highest. (d) Zoom-in on a pair of atoms for a few inter-atomic distances $R$. (e) Raman sideband spectra, averaged over all atoms, for two different orientations of the Raman beams (orange: $x$ and $y$, black: $y$ and $z$) showing all three modes of atomic motion (dashed lines). From the asymmetry of the sidebands we extract the mean motional number for each mode $\bar{n}_{x,y,z} = (0.11,0.11,0.56)$.}
\end{figure}

%\paragraph{Experimental setup: tweezers}
The experiment starts by trapping single $^{87}$Rb atoms in a two-dimensional array of tweezers. The array is generated by imprinting a phase hologram with a spatial light modulator (SLM) on a 810 nm laser beam which is then focused with a high-NA objective lens shown in Fig.~\ref{fig1}(b). The array consists of pairs of traps with an adjustable spacing ranging from 1.5 $\mu$m to 5 $\mu$m [Fig.~\ref{fig1}(c,d)], obtained by summing two holograms: one computed by the weighted Gerchberg-Saxton algorithm to generate a regular 2D array and a binary phase grating with tunable period to diffract each single trap into a pair. 
The measured trap depth is 0.62 mK and the trap frequencies are $(\omega_x, \omega_y, \omega_z) = 2\pi \times (147, 117, 35)$~kHz. From these values, we calculate a tweezers waist of 0.53/0.62 $\mu$m~\footnote{The tweezers radial anisotropy of 20~\% naturally emerges from vector diffraction theory predicting a 18~\% larger waist along the tweezers linear polarization ($y$) for the NA = 0.75 used here.} and a Rayleigh length of 1.56 $\mu$m, within 15~\% of the diffraction theory limit. Single atoms are randomly loaded in the traps and detected by taking $10$~ms fluorescence images at the beginning and end of each experiment. We post-select the runs where one or two traps of a given pair are loaded when necessary.

%\paragraph{Experimental setup: cooling}
By operating in a regime where the dynamics is driven by the dipole-dipole interaction $J = C_3/R^3$  (in contrast to the blockade regime where it is driven by the laser coupling $\Omega_{\rm cw}$), it becomes crucial to control the inter-atomic distance $R$ and to minimize its uncertainty $\Delta R$ as it translates into an interaction noise $\Delta J/J = 3 \Delta R/R$. 
Usually, experiments with Rydberg atoms in tweezers are performed with ``hot'' atoms ($T \sim 30 \, \mu$K) whose position fluctuates by $\sqrt{k_B T / m \omega^2} \sim 100 $~nm ($500$~nm along $z$), either because they rely on Rydberg blockade or because large distances ($R \sim 10 \, \mu$m) are used~\cite{deLeseleuc2020,Lienhard2020,Jaewook2020}. For us, these thermal fluctuations give an unacceptable noise $\Delta R_{\rm th}/R \sim 10 \, \%$. 
Here, we successfully suppress them by applying Raman sideband cooling to bring the atoms into the motional ground-state of the tweezers~\cite{Kaufman2012,Thompson2013}. From the sideband spectra shown in Fig.~\ref{fig1}(e), we extract the mean motional quanta $\bar{n} = (0.11, 0.11, 0.56)$ giving a position spread $\sqrt{\bar{n}+1/2} \sqrt{\hbar/m\omega}= (22,25,60)$~nm. This translates into an uncertainty $\Delta R_{\rm qu} = 35$~nm ($\Delta R_{\rm qu}/R < 2\, \%$) dominated at $90 \,\%$ by zero-point quantum fluctuations. We emphasize that, in this limit, the coupling with the external {d.o.f.}~is mostly coherent.

\begin{figure}
	\includegraphics[width=\linewidth]{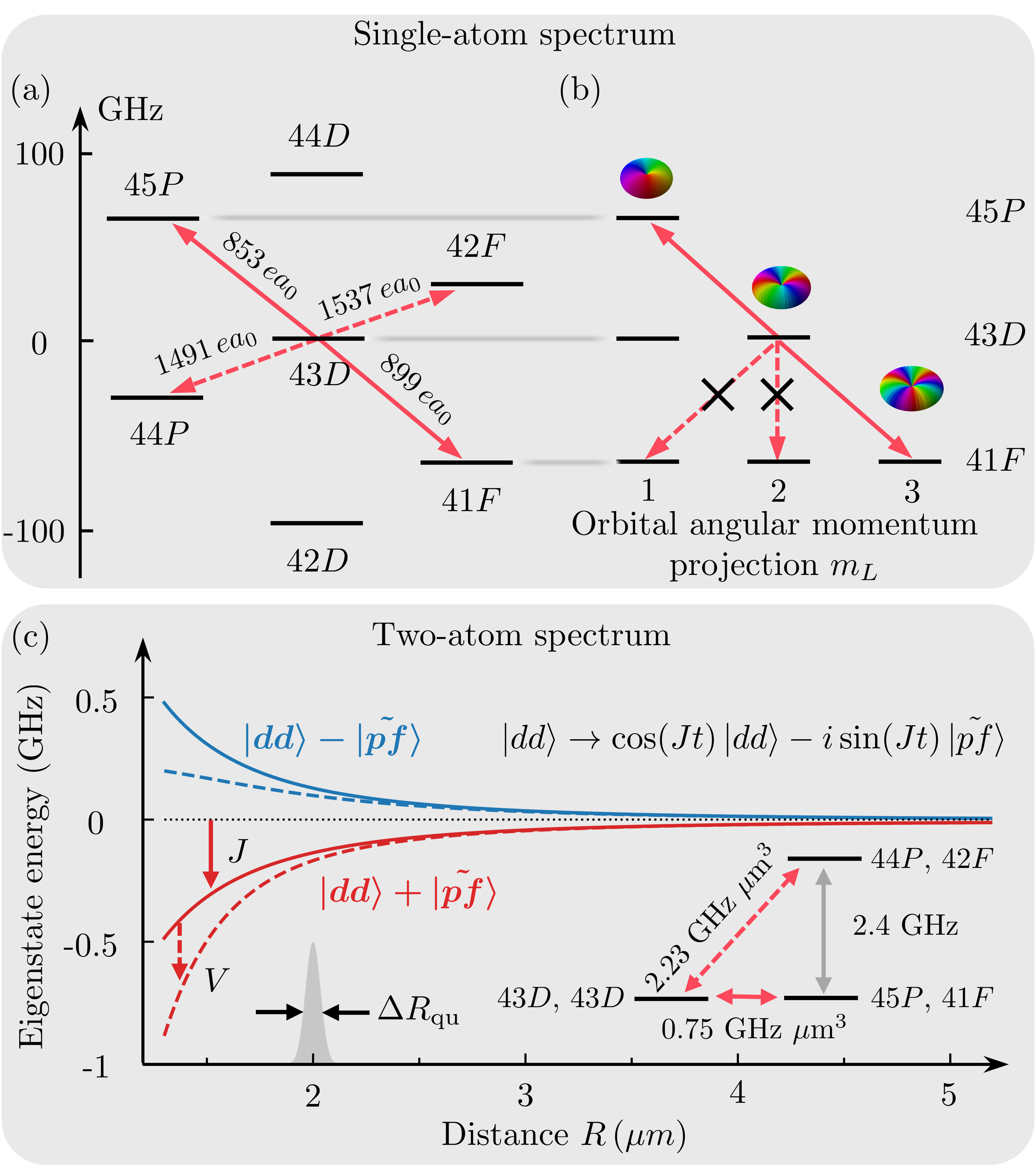}
	\caption{\label{fig2} \textbf{Rydberg levels and dipole-dipole couplings.} (a,b) Single-atom spectrum showing dipole matrix elements $d$ from state $43D$ (a), and the different $m_L$ sublevels (b), highlighting the selection rule for a pair of atoms aligned with the quantization axis. Atomic orbitals for $m_L =L$ are represented. (c) Two-atom spectrum obtained by diagonalizing $\hat{H}_{\rm dip}$. The inset shows the coupling $|C_3| = d_1 d_2 / 4\pi \epsilon_0$ and energy difference $\Delta E$ between the relevant pair-states. The resonant coupling gives a splitting $J = \sqrt{2}C_3/R^3$ between the two eigenstates (solid lines), while the off-resonant channel adds a van der Waals shift $V = -(\sqrt{2}C_3/R^3)^2/\Delta E$ (dashed lines). The probability distribution of inter-atomic distance $|\psi(R)|^2$ is shown for the measured $\Delta R_{\rm qu} = 35$~nm.}
\end{figure}

%\paragraph{Experimental setup: ultrafast excitation}
We now briefly describe the ultrafast ($\sim 100$ ps) coherent excitation of Rydberg atoms~\cite{Takei2016,Mizoguchi2020} with two single-photon Rabi $\pi$-pulses from the ground state $\ket{g} = \ket{5S_{1/2}}$ to the intermediate state $\ket{e} = \ket{5P_{3/2}}$, and then up to the Rydberg state $\ket{d} = \ket{43D_{5/2}}$, as depicted in Fig.~\ref{fig1}(b). A first 780~nm laser pulse (duration: 2~ps, bandwidth: 700~GHz) drives the $g \leftrightarrow e$ transition with its energy set to achieve a Rabi $\pi$-pulse, thus fully transferring the valence electron to the intermediate level. After tens of picoseconds, a 480~nm pulse drives the $e \leftrightarrow d$ transition. To account for the finite energy splitting of the Rydberg series [see Fig.~\ref{fig2}(a)], we cut the 480~nm bandwidth down to 100~GHz (14~ps) to avoid exciting the nearby $42D$ and $44D$ levels, as checked by numerical simulations. To achieve sufficient driving strength we focus the 480~nm beam to a diameter of $20 \, \mu$m; consequently only a single column of the array, shown in Fig.~\ref{fig1}(c), experiences the maximum intensity. We succeed in driving more than $ 2 \pi$ cycles on the $e \leftrightarrow d$ transition with a single pulse, or two $\pi$-pulses in a pump-probe configuration. Large pulse-to-pulse energy fluctuation of the 480~nm laser currently limits the $\pi$-pulse fidelity to $75 \, \%$. 

The excitation is performed in $\sim 100$~ps (dominated by the delay between the two pulses), fast enough to neglect spontaneous emission from the $5P$ orbital (25~ns),
%(which, given its 25~ns lifetime, is metastable on this timescale)
as well as interaction between Rydberg states. We also ensure that the electronic ($m_S$) and nuclear ($m_I$) spin degrees of freedom do not mix with the electronic orbital ($m_L$) after the excitation. To this end, we initially prepare the atoms in the spin-polarized state $\ket{5S_{1/2},F=2, m_F = 2} = \ket{5S,m_L = 0} \ket{m_S = 1/2, m_I = 3/2}$ and adjust the two excitation lasers to be $\sigma^+$-polarized to prepare the state $\ket{d} = \ket{43D, m_L = 2}\ket{m_S = 1/2, m_I = 3/2}$ (this also forbids the excitation of $S$-orbitals). From now on, we drop the spin degrees of freedom as they will not mix with the Rydberg orbital dynamics.

%\paragraph{Ultrafast interaction discussion}
\section{Ultrafast F\"orster oscillation}

Following the excitation, a pair of atoms in $\ket{dd}$ experiences the effect of the dipole-dipole Hamiltonian $\hat{H}_{\rm dip}$, which is dominated by the coupling to the pair state $\ket{pf} = \ket{45P;41F}$, as shown in Fig.~\ref{fig2}. The detuning $\Delta E$ of this channel is accidentally small in $^{87}$Rb ($-8$~MHz), giving rise to the so-called F\"orster oscillation between $\ket{dd}$ and the quasi-resonant symmetric state $\ket{\tilde{pf}} = \left(\ket{pf} + \ket{fp}\right)/\sqrt{2}$ with a coupling $J = \sqrt{2} C_3 / R^3$. With increasing interaction time $t$, the atoms evolve into $|\Psi(t)\rangle = \cos(Jt)\ket{dd} - i \sin(Jt)\ket{\tilde{pf}}$, and at $t = \pi/J$, they are back into $-\ket{dd}$ having acquired a $\pi$-phase shift. A refined description of the system requires to account for the finite size $\Delta R_{\rm qu}$ of the wavefunction $\psi(R)$ describing the relative position of the two atoms, leading to the following entangled state of internal and external d.o.f.: 

%\begin{equation}
\begin{align} \label{eq1}
|\Psi(t)\rangle = \int {\rm d}R  \,\psi(R) &\left[ \cos\left(\frac{\sqrt{2}C_3}{R^3}t\right)|R;dd\rangle \right. \nonumber  \\
&\left.- i\sin\left(\frac{\sqrt{2}C_3}{R^3}t\right)|R;\tilde{pf}\rangle\right]
\end{align}
%\end{equation} 

Other channels or degrees of freedom could perturb this ideal dynamics. We find that $\ket{44P;42F}$ is the next relevant channel, however for $R > 1.5 \, \mu$m the coupling remains weaker than the energy difference, only giving a weak non-resonant perturbation, i.e., a van der Waals shift $V$ [Fig.~\ref{fig2}(c)]. Other dipole-dipole channels, as well as higher-order ones (e.g., dipole-quadrupole) are further negligible: they are taken into account in numerical simulations but are left aside of the discussion. 
We now consider the projection of orbital angular momentum $m_L$, whose dynamics is in general non-trivial as $\hat{H}_{\rm dip}$ contains terms changing each atom angular momentum~\cite{Ravets2015} by $\Delta m_L = 0,\pm1$. As shown in Fig.~\ref{fig2}(b), with our choice of initial state and by aligning the pair of atoms with the quantization axis such that the total angular momentum projection is conserved, we ensure that $m_L = L$ throughout the dynamics which effectively decouples this degree of freedom.

%\paragraph{Main result 1: Forster oscillation}

 \begin{figure}
	\includegraphics{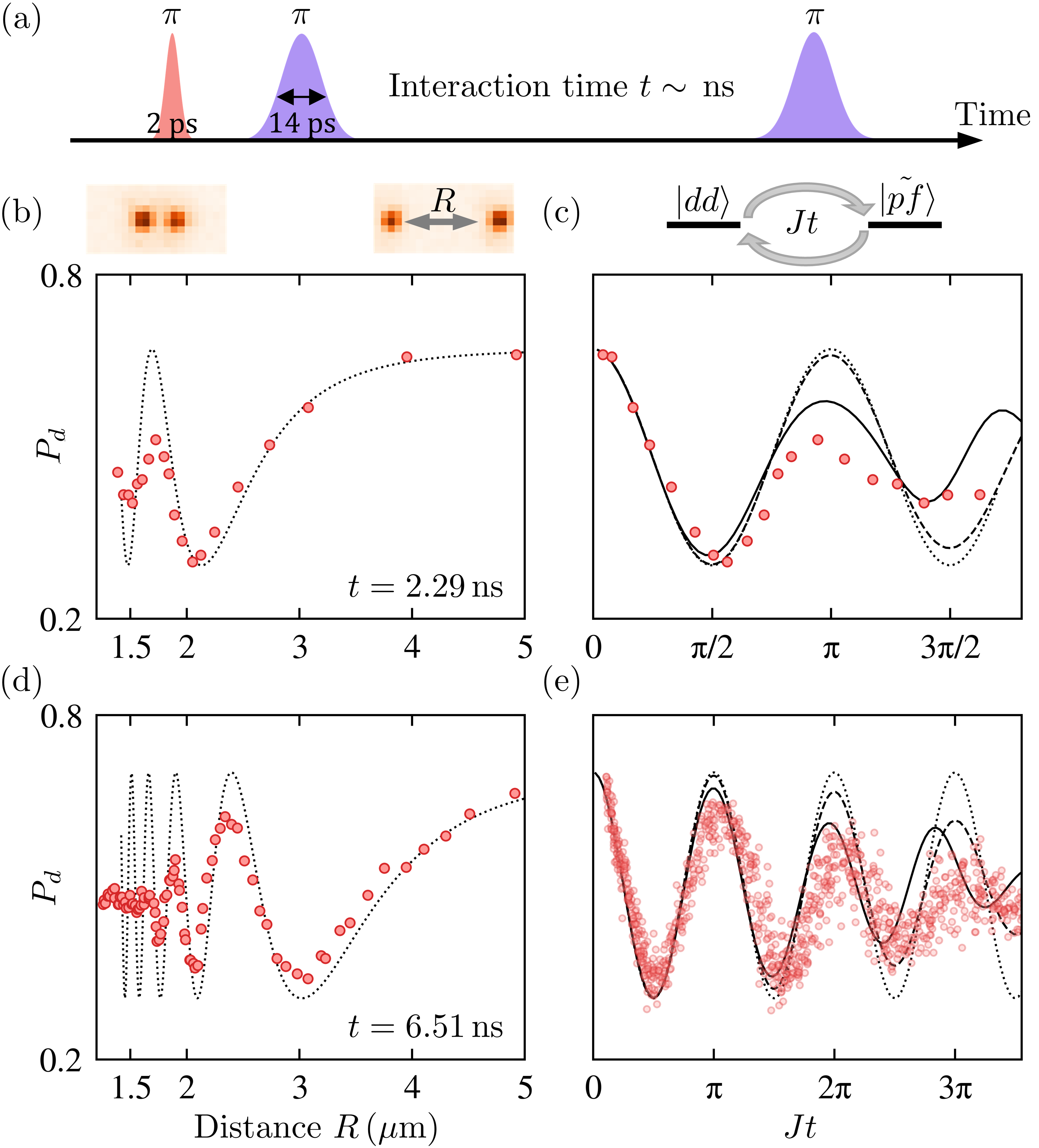}
	\caption{\label{fig3} \textbf{Ultrafast F\"orster oscillation.} (a) Pump-probe sequence. (b-e) Probability $P_{d}$ to be in state $\ket{d}$ after a delay $t =$ 2.29 ns (b,c) or $t =$ 6.51 ns (c,e). The oscillation is shown as a function of the interatomic distance $R$ (b,d), or the interaction area $Jt = \sqrt{2}C_3t/R^3$ (c,f). The black curves are simulations of the ideal F\"orster oscillation rescaled by the SPAM errors (dotted), including the effect of quantum uncertainty in position (dashed), and perturbing interaction channels (solid). In (e), we show the data of all 16 pairs of atoms, while this is averaged in (b-d). For each distance, we repeat the experiment 2000 times, out of which we select the $\sim 25 \, \%$ shots where both atoms of a pair are loaded, resulting in a quantum projection noise $< 3 \, \%$ for each point in (e), and $< 1 \, \%$ in (b-d).}
\end{figure}

To observe the ultrafast F\"orster oscillation, we first perform a pump-probe experiment, as illustrated in Fig.~\ref{fig3}(a). 
A first 480 nm $\pi$-pulse brings the atoms to $\ket{d}$ and initiates the interaction-driven dynamics. After a delay $t$ of a few nanoseconds, a second $\pi$-pulse de-excites the atoms in $\ket{d}$ to $\ket{e}$, which are then recaptured in the tweezers, while atoms in other Rydberg states are field-ionized.   
Figure~\ref{fig3}(b,d) show the probability for atoms to be detected in $\ket{d}$ after a delay $t = 2.29$~ns or $t = 6.51$~ns as a function of the measured distance $R$~\cite{measured_R}, which is varied by adjusting the binary phase grating. We also display the oscillation as a function of the interaction area $J t = \sqrt{2} C_3 t / R^3$ [Fig.~\ref{fig3}(c,e)]. 
Satisfyingly, the period matches the ab-initio theory (dotted curves) for the two datasets, clearly demonstrating that the dynamics is coherently driven by the F\"orster channel.

We now discuss the finite contrast and damping of the oscillation. The 40~\% contrast is caused by state preparation and measurement (SPAM) errors, i.e., imperfect $\pi$-pulse in the excitation and de-excitation processes, and is not related to the F\"orster oscillation. These errors are independently measured and used to rescale the contrast of all simulations, giving good agreement with the experimental data.
Concerning the damping, a first source is the quantum uncertainty of the atoms' position in their traps. Calculations (dashed curves), consisting in tracing over the external d.o.f. in Eq.~(\ref{eq1}), demonstrate that this effect is indeed responsible for most of the damping for large values of $Jt$, as clearly seen in Fig.~\ref{fig3}(e).
The effect of perturbing channels (solid lines) effectively gives an additional damping, which is more pronounced in Fig.~\ref{fig3}(c), where we use smaller distances increasing the perturbation.
%Promisingly, for a both effects are barely visible at $Jt = \pi$, these two effects plays a larger role for shorter distance $R$ (and thus time $t$) as clearly seen when comparing Fig.~\ref{fig3}(c,e). 
By plotting data points for each pair of trap, we also notice a horizontal scatter increasing with $Jt$. These errors in the measured distance $R$ are most likely caused by our assumption that atoms in a pair are exactly in the same $z$-plane~\cite{measured_R}. 
Eventually, there remains a slight discrepancy between the measured damping and our simulations. We attribute this to imperfect polarization of the excitation lasers leading to a coupling of the ideal F\"orster dynamics with the $m_L$ and $m_S$ degrees of freedom. These last two points could be addressed in future experiments by devising independent in-situ measurements. 

%Concerning the damping, it is first interesting to compare Fig.~3(c,e) showing that, for a fixed interaction area $Jt$, the damping is stronger for shorter distance $R$. 
%A first source of damping could be the coupling of the $m_L$ degree of freedom (caused by imperfect preparation $m_L < L$ or bad alignment of the pair of atoms), but this effect only depends on $Jt$ and cannot explain the observation made above. 
%A second source is the position fluctuation of the atoms in their tweezers and this effect is stronger for smaller distance $R$. Figure~3(c,e) shows simulations for atoms before ($T = 50 \, \mu$K) and after Raman sideband cooling (dashed and dotted). The observed damping is clearly slower than for $T = 50 \, \mu$K atoms for t = 6.5 ns, but faster for t = 2.3 ns, and this inconsistency rules out the position fluctuation as the main source of damping in Fig.3(c). 
%A third effect is the spread of tweezers separation over the 14 pairs of the column. CHECK THIS.
%A fourth effect could be the coupling to the $(n+1)P(n-1)F$ channel, shown in Fig.~2(b), which cannot be neglected when $R$ decreases towards $1.3 \, \mu$m. However, numerical simulations also rules out this effect as it only creates a $<10$~\% damping at $Jt = 2\pi \times 0.5$ for $t = 2.3$ ns, much smaller than the observed one. 
%A fifth effect could be the presence of an electric field coupling the $F$ state to hydrogen-like orbitals ($L > 3$), but it would only gives a damping proportional to $Jt$...

%\paragraph{Main result 2: Entanglement resource}
\section{Conditional phase}

\begin{figure}
\includegraphics{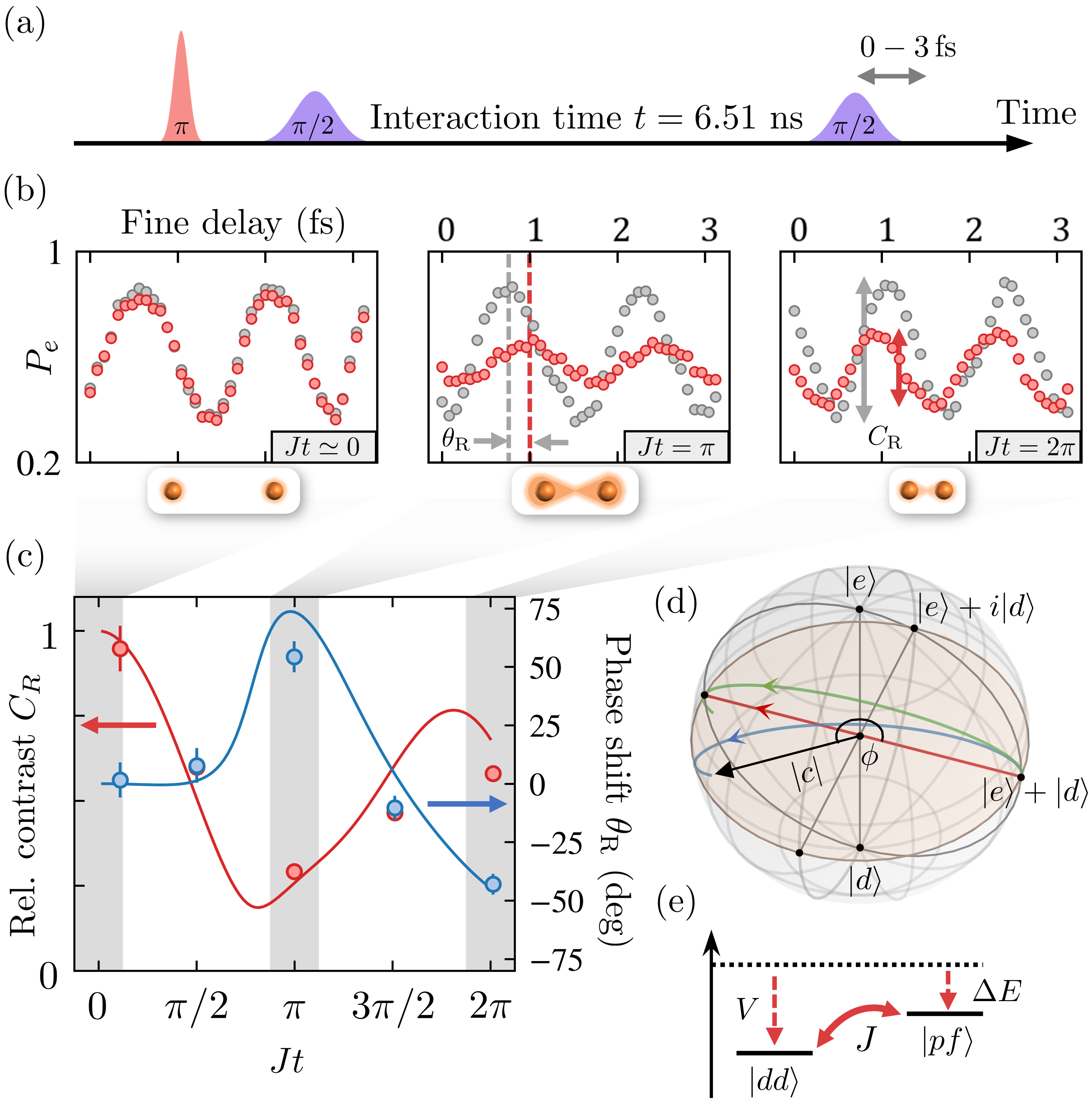}
\caption{\label{fig4} \textbf{Probing the conditional phase $\phi$.} (a) Ramsey pump-probe sequence. (b) Ramsey fringes for $Jt = 0.1\,\pi, \pi, 2\pi$ ($R = 5, 2.4, 1.9 \, \mu{m}$) when one (gray) or two (red) atoms are loaded in a pair of tweezers. The fine delay offset is arbitrary defined to be 0~fs in each interferogram. (c) Relative Ramsey contrast $C_R$ and phase shift $\theta_R$ extracted from (b) with numerical simulations. Errors bars represent the estimated fitting error. (d) Bloch sphere representation of the state of atom 1 which remains in $\ket{e}+\ket{d}$ when atom 2 is in $\ket{e}$, but is driven into $\ket{e}+|c|e^{i\phi}\ket{d}$ when atom 2 is in $\ket{d}$. Trajectories of the state vector are shown for three scenarios depending on the parameters $(J,V,\Delta E)$ illustrated in (e): (red) resonant energy-exchange $V = \Delta E = 0$, (blue) real parameters $(J,V, \Delta E)/(2\pi) = (75, -23, -8)$~MHz at $R = 2.4 \, \mu$m, and (green) with envisioned active correction $\Delta E = -V$.}
\end{figure}

We then investigate the conditional phase $\phi$ imprinted on a pair of atoms. To this end, we now consider atoms initialized in a coherent superposition $\ket{e} + \ket{d}$ of the intermediate and Rydberg state. The interaction drives the state of atom 1 of a pair into $\ket{e} + |c|e^{i\phi} \ket{d}$ when atom 2 in state $\ket{d}$. Ideally, $|c|e^{i\phi} = \cos(Jt)$ would realize a controlled-Z gate at $Jt = \pi$.
Having previously measured the population $|c|^2$, we now gain sensitivity to the phase $\phi$ by performing a Ramsey pump-probe experiment~\cite{Takei2016,Chunmei2018}. As depicted in Fig.~\ref{fig4}(a), a first $\pi/2$-pulse prepares the coherent state, which is then probed after an interaction time $t = 6.51$~ns with a second $\pi/2$-pulse whose arrival time is scanned with attosecond precision over 3 fs to realize a variable phase $\theta$.
%~\footnote{The phase of this pulse $\theta/(2\pi) = t/T$ (in the rotating frame of the atomic coherence) is scanned by adjusting the delay to record a Ramsey fringe with a period $T = h/(E_d - E_e) = 1.6$~fs.}.
The Ramsey signal given by atom 1, conditioned on the state of atom 2, could be observed with single-atom addressing techniques~\cite{Saffman2019,Hennrich2020}. Here, Ramsey fringes are recorded irrespective of the state of atom 2 ~\cite{Mandel2003,Jaewook2020}, thus giving a signal oscillating as $\cos(\theta) + |c|\cos(\theta+\phi) = C_R \cos(\theta + \theta_R)$ with a relative contrast $C_{\rm R}$ and phase shift $\theta_{\rm R}$ from reference fringes obtained with a single atom.
In the ideal case, the interaction produces a maximally-entangled state at $Jt = \pi$ ($C_R = 0$), and returns the atoms to a pure product state at $Jt = 2\pi$ ($C_R = 1$, $\theta_R = 0$). While the measured fringes shown in Fig.~\ref{fig4}(b) approximately follow this simple scenario, there are clear deviations: the contrast is imperfect and there are phase shifts.

%: the contrast is neither perfectly canceled ($Jt = \pi$), nor revived ($Jt = 2\pi$) and there are clear phase shifts $\theta_{\rm R}$ in both case.

To capture these, we need to account for the van der Waals shift $V$ caused by the off-resonant channels, and the natural energy defect $\Delta E$ of the resonance. In the limit $V, \Delta E \ll J$, we can derive $|c|e^{i\phi} \simeq \cos(Jt)e^{-i(V+\Delta E)t/2}$, which is represented on the Bloch sphere. We then show, in Fig.~\ref{fig4}(c), the expected Ramsey contrast and phase shift as a function of the interaction area $Jt$ as two atoms are brought closer. We now obtain an excellent agreement with the measurements, supporting that we accurately capture the evolution of the conditional phase $\phi$.
%first relate the Ramsey signals $C_{\rm R}$ and $\phi_{\rm R}$ to the interaction phase $|c|e^{i\phi}$ (see methods), which for the limiting case $|c|= 1$ (valid when $Jt = n \pi$) gives $C_{\rm R} = |\cos(\phi/2)|$ and $\theta_{\rm R} = \phi/2$. 
%Combining these two results, we show how the expected Ramsey signals in Fig.~\ref{fig4}(e) varies with the interaction area $Jt$ as two atoms are brought closer. We now obtain a very good agreement with the measurements, highlighting the importance of considering the additional van der Waals shift $V$. 
%Figure~4(e) shows how the relative contrast $C_{\rm R}$ and phase shift $\theta_{\rm R}$ of the Ramsey fringe varies 
For our experimental parameters, we calculate that $\phi = 1.17 \, \pi$, thus overshooting the ideal $\pi$-phase shift. 
This could be corrected by tuning the energy of the resonant $\ket{pf}$ state to $\Delta E = -V$, for example with a weak electric field ($\rm < 1 \, {\rm V/cm}$) applying a Stark shift~\cite{Ravets2014}. 
%For $Jt = \pi$, the atoms are expected to be in agreement with the observation of a minimal fringe amplitude. For larger $Jt$, the contrast increases again as the atoms de-entangle.
%We attribute the imperfect revival at $Jt = 2\pi$ to effects described in the previous paragraph, and the remaining contrast at $Jt = \pi$ to laser power fluctuations, giving rise to shot-to-shot deviations from ideal $\pi/2$-pulses creating imperfectly balanced superposition.
Finally, we point that a related interaction-driven procedure was used by~\citet{Jaewook2020} to obtain a conditional phase between Rydberg atoms at a distance $R \sim 10 \, \mu$m, using a pure van der Waals coupling, in a time $\pi/V = 2.6 \, \mu$s. Here, we demonstrate a 400~times faster dynamics thanks to the use of ultrafast pulsed-laser technology to excite Rydberg states in tens of picoseconds, and much closer distances bringing the interaction strength towards the GHz range. 

\section{Discussion and conclusion}
%Discussion
The next step towards an ultrafast $C_Z$ gate would be to improve the fidelity of the pulsed excitation; and to map a long-lived hyperfine qubit to a ground-Rydberg superposition, which can also be achieved in an ultrafast timescale set by the 6.8~GHz ground-state hyperfine splitting~\cite{Monroe2010}. Errors originating from perturbing channels, such as leaks in other pair-states $|r'r''\rangle$ or dynamical phases, can be eliminated by fine-tuning the energy levels using electric fields (dc-Stark shift) or non-resonant laser or microwave fields (ac-Stark shift), to achieve synchronization of maximal entanglement and minima of leaks, as performed with SC qubits~\cite{GoogleSC2019,DiCarlo2019,GoogleSC2020,DiCarlo2021}.
The fundamental source of infidelity for an interaction-driven gate is entanglement with the external d.o.f.~of the atoms, with the quantum uncertainty in distance leading here to a finite probability for the atoms to remain in the resonant $\ket{pf}$ state of $|\sin(Jt)|^2 \simeq (3 \pi \Delta R/R)^2 = 1.8 \, \%$. 
%Leakage in non-resonant channels are mostly insensitive to $R$ and can be suppressed by 
%While we have demonstrated parameters where the effect of quantum uncertainty in distance and additional channels are barely noticeable at the speed limit $t = \pi/J$ [cf. Fig.~\ref{fig3}(e)], 
We envision that advanced coherent control techniques can further suppress such errors. The motional coupling can be reduced by an order of magnitude by preparing squeezed states of motions~\cite{Salomon1999,Burd2019}, and even further by using a decoupling echo-like sequence~\cite{Zoller2006}. One could also dynamically adjust the F\"orster resonance defect $\Delta E$ to perform robust fast adiabatic gates~\cite{Martinis2014}.

In summary, we have observed an ultrafast quasi-resonant energy exchange between two single Rydberg atoms trapped in individual tweezers, with an inter-atomic distance controlled with a precision limited by quantum fluctuations. The interaction-driven dynamics gives rise to a calculated conditional phase $\phi = 1.17 \, \pi$, in good agreement with observations. This is the key resource for an ultrafast two-qubit gate operating at the fundamental speed-limit set by the dipole-dipole Hamiltonian, and not by technical limitations set by the available power of cw lasers. Such a gate would be two orders of magnitude faster than with state-of-the-art protocols based on the blockade mechanism, thus evading decoherence mechanisms affecting Rydberg atoms on a timescale of tens of microseconds.  
%The entanglement dynamics demonstrated here is two orders of magnitude faster than in protocols based on the Rydberg blockade mechanism, and saturates the speed limit $t_J = \pi/J$ set by the dipole-dipole interaction on the chosen channel, which is promising for suppressing sources of decoherence affecting Rydberg atoms.
Furthermore, our experimental platform could be used to simulate the dynamics of interacting spin-1 models by bringing more than two atoms at close distance, to investigate Rydberg interactions when orbitals of neighbor atoms overlap~\cite{Mizoguchi2020}, or the use of Rydberg wavepackets instead of single Rydberg levels.
%This demonstration opens for ultrafast quantum simulation and computation.

\section*{Acknowledgments}
We acknowledge D. Barredo, A. Browaeys, D. Jacksch, T. Lahaye and M. Weidem\"uller for fruitful discussions about this work.  We thank Z. Meng, V. Bharti and N. Takei for contributions at the early stage of this work; R. Villela for a careful reading of the manuscript; Y. Okano, T. Toyoda, M. Aoyama and H. Chiba for the technical support; R. Iino and R. Yokokawa for fruitful discussions on the development of the objective lens and for providing equipment to evaluate its performance; and, from Hamamatsu Photonics K.K., Y. Ohbayashi for coating the viewports and H. Toyoda, Y. Ohtake, T. Ando, H. Sakai, Y. Takiguchi, K. Nishimura for fruitful discussions about the optical system and providing the spatial light modulator.
This work was supported by MEXT Quantum Leap Flagship Program (MEXT Q-LEAP) JPMXS0118069021 and JSPS Grant-in-Aid for Specially Promoted Research Grant No. 16H06289. T.T. and S.dL. acknowledge partial support by JSPS Grant-in-Aid for Research Activity Start-up (19K23431, and 19K23429 respectively). K.O. acknowledges partial support by the Alexander von Humboldt Foundation and Heidelberg University.

\section*{Author contributions}
Y.C., T.T., T.P.M. and S.dL. designed and carried out the experiments. Y.C. and S.dL. performed data analysis, theory and simulations, and wrote the manuscript with contributions from all authors. S.dL. and K.O. supervised and guided this work.

%\section*{Data availability}
%The data supporting the findings of this study are available from the corresponding author upon reasonable request.

%\section{Competing interests}
%The authors declare no competing interests.

\bibliographystyle{apsrev4-2}
\bibliography{abbreviated}

\begin{thebibliography}{52}
\bibitem [{mea()}]{measured_R}%
  \BibitemOpen
  \href@noop {} {}\bibinfo {note} {The distance $R$ is extracted from
  fluorescence images giving the $x,y$ position of each atom. The magnification
  of the imaging system was calibrated to 1~\% using a retro-reflected resonant
  beam as an in-situ ruler, giving a scattering rate spatially modulated with a
  period of 390~nm which we detect by scanning the position of the
  atoms.}\BibitemShut {Stop}%
\end{thebibliography}

\end{document}